\documentstyle[prl,aps,twocolumn]{revtex}

\begin{document}
\title{ 
Stochastic Resonance in Maps and  Coupled Map Lattices.}
\author{Prashant M. Gade}
\address{Jawaharlal Nehru Centre for Advanced Scientific Research \\
Jakkur, Banglore-560064,  INDIA}
\author{ Renuka Rai and Harjinder Singh}
\address{
 Department of Chemistry,\\ Panjab University, Chandigarh-160014, INDIA}

\maketitle
\begin{abstract}
We demonstrate the phenomenon of
stochastic resonance (SR) for discrete-time dynamical systems. We 
investigate various systems that are not necessarily bistable, but do have 
two well defined states, switching between which is aided by external  
noise which can be additive or multiplicative.
Thus we find it to be a fairly generic 
phenomenon. In these systems, we investigate kinetic aspects like
hysteresis  which reflect the 
nonlinear and dissipative nature of the response of the system to the external 
field. 
We also explore spatially extended systems with additive or parametric noise
and find that they differ qualitatively.
\end{abstract}

\section{Introduction}

A seemingly counterintuitive scenario that a weak signal can
be enhanced by addition of noise was proposed by Benzi and
coworkers \cite{Ben} in connection with the glaciation cycle
of the earth. Since then stochastic resonance(SR) has been employed
in explaining various phenomena (see e.g. \cite{Nat},\cite{Stat})
and has been studied extensively experimentally as well as theoretically
\cite{McW,Sct,Roy}. The major emphasis 
of the studies has been on the original model 
by Benzi {\it et al} \cite{Ben} in which the
stochastic system in consideration has
two stable fixed points in absence of noise and the driving force,
though few other models have also received 
attention \cite{Euro,fn,Sto,kw}. 
The essence of the phenomenon is that even a weak periodic signal which
is undetectable in absence of noise can force a bistable system to switch
between its two states, periodically, in presence of an optimal noise.
Often  one calculates the power spectrum of the output
signal of the system filtered through a two-level filter. The ratio
of output power in the frequency of the signal with the background noise,
also called as signal to noise ratio ($SNR$) is a relevent quantifier here.
In fact, nonmonotonic behavior of $SNR$ with the noise intensity has 
become a 'fingerprint' of this phenomenon.

Additive noise has been the focus
of most of the studies. The multiplicative noise, which is
not equivalent to additive noise in the presence of a periodic
field \cite{Risken} and thus in principle can show
qualitatively different behavior, has not been studied much \cite{Gam}. 
Multiplicative noise occurs in
a variety of  physical phenomena\cite{Rei,mul,Gam} and is certainly important.
Also of importance are the kinetic aspects of the phenomenon which 
reflect the way a system responds to the signal. 
These aspects have been mainly investigated by measuring the phase
shift between the signal and the response \cite{Gamm,rest}. 
Hysteresis which reflects
phase shift and also the extent of losses in a periodically
driven  system, is an equally
important quantifier \cite{She,RKP,DT}.
However, studies of usual hysteresis behavior in presence of noise have
hardly been attempted. One notable exception has been a numerical work by 
Mahato and Shenoy\cite{She}. However, they define the hysteresis loop
in a different way than is usually  done.  Thus, as they state, we cannot
expect their results to be carried over  to the usual case even
qualitatively.
Recently there have been few interesting studies on SR in spatially 
extended systems \cite{Lind} (see also \cite{ben85,Jung92}). The
reasons for the recent surge of interest in analysis of 
spatiotemporal systems are
not far to seek. These
systems are important from the point of view of potential applications
that range from coupled nonlinear devices and signal processing to
neurophysiology  and  merit further attention.

We feel that SR is a generic feature of two-state systems
neither state of which needs to  be a stable fixed point. 
Any system with two well defined and well separated states 
switching between which can be aided by noise, can possibly show 
stochastic resonance in the presence of a weak signal. We will illustrate 
this with  systems switching between two chaotic attractors, a chaotic 
attractor and  a fixed point, and will also present the standard model 
of two stable fixed points. It is easy  and  computationally
inexpensive to construct such cases using discrete-time 
systems like maps and we will be using maps for demonstration.
We have used both additive and multiplicative noise in our simulations.
In many physical and chemical systems noise is generated internally
(see e.g. \cite{Rei}) and such systems have been observed to show stochastic 
resonance.  As we have pointed out, the multiplicative noise in the presence
of a periodic force is not equivalent to additive noise coupled with a periodic 
signal. Interestingly, in the case of a single map,
both show a similar qualitative 
behavior as far as SR is concerned, i.e., in both cases $SNR$ shows 
nonmonotonic behavior of response as a function of noise and has
a peak at some value of noise. 

In this dissipative and nonlinear
system, the response to external field is likely to be delayed
and nonlinear. The simplest way to gauge it is to study hysteresis
in these systems which will reflect both losses as well as delay in the 
response \cite{She}. 
This will be information additional to the one given by simple signal 
to  noise ratio, e.g., $SNR$ does not reflect the phase shift in the
response.  

Finally we investigate spatially
extended systems where local dynamics is governed by these maps.
Such systems, popularly known as coupled map lattices 
(CML)\cite{kan} have gained considerable attention in recent times
due to their computational simplicity, and ability to reproduce
qualitative features in various phenomena. To name a few interesting
applications one can point out the modelling of phase ordering 
dynamics \cite{Oono}, spatiotemporal 
intemittency\cite{Spati}, spiral waves \cite{spi} etc.. 
However, we are not aware of any attempt to observe SR in these systems. 
Here we see a qualitative difference between the systems 
subjected to an additive noise and parametric noise.
 
The section II and III we will define our models and present their
analysis. In section IV we will discuss the hysteresis
and effect of noise. In section V we will define the
spatially extended systems in a way popularly
known as coupled map lattice and discuss results in it.
Finally in section V we conclude and discuss questions that
we are interested in.

\section{The Models}

Let us consider the following maps in absence of any periodic or noisy
drive.	
\begin{eqnarray}
f_S(x)&=& S \; tanh(x)\;\; \; x \in (-\infty,\infty)\; S>0 \label{1}\\
g_a(x)&=& \exp (a) x \vert_{\rm mod 1}\;\; \; x \in (-1,1) \label{2}\\
h_r(x)&=& r x \vert_{\rm mod 1}\;\; \;x \in (0,{\frac{1}{2}}] \nonumber \\
&=& r (1-x)\vert_{\rm mod 1}\;\; \;x \in({\frac{1}{2}},1] \nonumber \\
h_r(x)&=&-h_r(-x)\; \; \; x \in (-1,0)\label{3}
\end{eqnarray}	
\noindent
a)The map $f_S(x)$ which has a hamiltonian symmetry ($f_S(-x)=-f_S(x)$) and 
two stable fixed points symmetric around $x=0$, is  clearly analogous
to the original model of bistable potential\cite{Ben}. In this case also, the
basin of attraction of positive fixed point $x_S^*$ is $(0,\infty)$ while
that of the negative fixed point that is symmetrically placed at
$-x_S^*$ is $(-\infty,0)$. Thus the system has 
two stable fixed point attractors any of which is reached depending 
on initial conditions. (See Fig. 1(a))\cite{fn2}

\noindent
b)The map $g_a(x)$ shows a chaotic behavior if $ a  >0$ or 
has a stable fixed point $0$ as attractor if $ a  <0$. Thus the 
map has a chaotic or fixed point attractor depending on value of 
$a$. (Fig. 1(b) shows the map $g_p(x)$ and $g_q(x)$ for $p>0$ 
and $q<0$.)

\noindent
c)The map $h_r$  also has a hamiltonian symmetry like map $f_S(x)$. 
It has a fixed point attractor at $x=0$ for $\vert r \vert <1$. 
However for $\vert r \vert >1$, it has an interesting behavior.
In this range, it shows two chaotic attractors symmetric around zero.
The map is such that for any initial condition $x_0 \in [0,1]$, 
$h^T(x_0) \in [0,1]$ for all times $T$.
In fact, for $1<r<2$ the attractor on the positive side
does not span an entire
unit interval but is in the interval $[a_0,a_1]=
[h^2({\frac{1}{2}}),h({\frac{1}{2}})]$ for $x\in (0,1)$, while 
for $x \in (-1,0)$, the values assymptotically span interval $[-a_1,-a_0]$.
(For $r\geq 2, \; a_0=0,\;a_1=1$)
This is a system with two symmetric chaotic attractors any of which is
reached depending on initial conditions. The attractors are separated by
$2 \;a_0$.  (See Fig. 1(c))

\section{SR in Maps}

The above maps have the desired property of having two
different attractors between which the system can switch 
when aided by noise.
We investigate the following systems.
\begin{eqnarray}
a) x_(t+1)&=& f_S(x(t)) +z cos(2 \pi \omega t)+ \eta_t \label{a)}\\
a1) x_(t+1)&=& f_{S+\eta_t}(x(t)) +z cos(2 \pi \omega t) \label{a1)}\\
b) x_(t+1)&=& (g_{a+\eta_t+z cos(2 \pi \omega t)} (x(t)-p_0))+p_0\vert_{\rm mod 1} 
\label{b)}\\
c) x_(t+1)&=& (h_{r}(x(t)) +z cos(2 \pi \omega t) +\eta_t)\vert_{\rm  mod 1} 
\label{c)}\\
c1) x_(t+1)&=& (h_{r+\eta_t}(x(t)) +z cos(2 \pi \omega t))\vert_{mod 1} 
\label{c1)}
\end{eqnarray}
where $\eta_t$ is delta correlated random 
number with variance $D$. Except in case  of eq.\ref{a)}, where $\eta_t$ has
a gaussian distribution, we have a uniform distribution for $\eta_t$.  
The maps $f_S(x)$ and $h_r(x)$ have hamiltonian symmetry, {\it i.e.}, 
$f_S(-x)=-f_s(x)$ and $h_r(-x)=-h_r(x)$. They  have
symmetric attractors on positive and negative side. Since we are 
interested in inter-state switching we neglect the intra-state 
fluctuations in cases of eqs. \ref{a)}, \ref{a1)}, \ref{c)} and \ref{c1)} 
while analysing the output.
The output is analysed in the usual manner, i.e. one takes the 
fourier transform of the time series thus generated and averages the 
power over various phases and also initial conditions. 
The $SNR$ is defined as the ratio of the intensity of $\delta$-spike
in the power spectrum at the frequency $\Omega=2\pi\omega$ to the
height of the smooth fluctuational background $Q^0(\Omega)$ at the 
same frequency $\Omega$ then
\begin{equation}
SNR =\log _{10}{\frac{{\rm Total} \;{\rm power}\;{\rm  in}\;{\rm the}\;
{\rm frequency }\;\Omega}{Q^0(\Omega)}}
\end{equation}
Variations in this definition do not change the 
results qualitatively.

a) This is  the simplest system which
mimics well studied 
bistable potential model \cite{Ben} of Benzi and coworkers.
Here the system toggles between the positive and the negative 
fixed points of the map $f_S$, ($x_S^*$ and $-x_S^*$). 
The system is defined over the
entire range $(-\infty, \infty)$. We apply gaussian noise.
It is seen that  as the noise intensity increases
the spectral strength of the signal also increases, but 
this happens at the
expense of noise, thus increasing signal to noise ratio. This happens 
since when the signal is at its peak, the little noise aids the system to 
flip from the basin of attraction of one fixed point to other. For 
large noise, the flips can occur almost all the time, the regularity 
is reduced and $SNR$ decreases again.
This behavior is shown in Fig. 2a) which shows $SNR$ as a function
of noise intensity $D$ for the system defined by eq.\ref{a)}

a1) In this map we also have a possibility of parametric noise
as in eq. \ref{a1)}. The value of noise changes the position 
and the stability of the fixed point. 
For large enough noise the fixed point can come arbitrarily
close to zero, which coupled with the periodic signal can cause flips
which can be very regular at optimal noise level. Thus $SNR$ shows
as standard SR behavior as a function of $D$ (See Fig. 2b).

b)  Now we explore the  possibility of competition
between a fixed point attractor and a chaotic
attractor switching between which is aided by noise.
(We have added a small constant
$p_0$ in the eq. \ref{b)} unlike eq. \ref{2} for numerical reasons.
The position of the fixed point now changes to $p_0$.
For $p_0=0$, trajectory that comes close to zero within numerical
precision will stay there.
This change does  alter the description of the map qualitatively.)
At the minimum value of the drive,
it is likely that
$ a_t= a + 
\eta_t + z \;cos(2 \pi\omega \;t)<0$ and the system will
be attracted to the fixed point. 
While small noise will aid this repetitive attraction towards fixed point,
very large noise is likely to reduce it. As a result, we see a nonmonotonic
response of $SNR$ to noise intensity. In some sense, this system mimics 
excitable dynamics,  where SR has been observed. \cite{kw} 
(though the excited state in this case is  not  a 
chaotic state.) Fig. 2c) shows $SNR$ as a function of noise intensity $D$
for this system.

This system is like a random walk when viewed on a logarithmic
scale. Neglecting the modulo factor, the variable value at time $n+1$ goes as
$ln(x_{n+1})=a_n+a_{n-1}+\ldots+a_0+ln(x_0)$ where $a_t
=a+\eta_t+z \cos(2 \pi \omega t)$.
Thus it is like a random walk
with initial position $ln(x_0)$ and  dispacement $a_i$ at
$i$th time step. The value of $x_i$ is bounded from above by
unity due to modulo condition and $x_i$ does not tend to zero
asymptotically since $a>0$. Due to $a>0$, this is
a case of a biased 1-d random walk bounded from above.
Thus this system is comparable with
stochastic resonance seen in random walk.\cite{PRR} Fig. 3 shows a
schematic diagram for the above description.

c) This is another  interesting but unexplored possibility.
Here  we have two chaotic attractors switching between which 
is aided by noise and a periodic signal. We would like to point 
out that apart from the nature of the attractors, this 
 system is very similar to 
system defined by eq.\ref{a)} as far
as the dynamics of interstate switching is concerned. Let us consider 
system  defined in eq.\ref{c)}. 
As noted above, the two attractors are well separated in 
absence of noise and periodic signal and the system stays in either of them 
depending on initial conditions.   
However, in the presence of noise, the  system can 'leak' out of the attractor.
This 'leaking', i.e. switching from positive attractor to negative and
vice versa is likely to occur at most opportunate times, i.e.,  at the
minimum and the maximum of the signal.  Thus
one may
expect stochastic resonance here which is indeed the case (See Fig. 2d).

c1) Here one could have a parametric noise as an alternative to additive noise.
The parametric noise can change the value of $a_0$ which controls
the distance between two attractors thus aiding the switching.
Fig. 2e show $SNR$ as a function of noise intensity $D$ in these 
systems. One can see a clear non-monotonicity in response. 

\section{Hysteresis}

Now we discuss the kinetic aspects of this phenomenon.  
 Hysteresis is a kinetic phenomenon which is the signature of
the response of the system to external field sweep. In general,
due to the frictional losses,  system is not able 
to follow the external signal exactly. There is an accumulated
strain after which it responds to the signal. The quantity that
reflects these losses is the area of the hysteresis loop. 
The most familiar example is the behaviour of magnetization $M$
as a function of alternating external magnetic field $H$.
We have a two-state system in our examples and we are  
analyzing the signal filtered through two state filter. This makes it
easy to define an analogue of magnetization. We define
\[m(t)={\frac{1}{N}}\sum_{j=1}^N sgn(x(jT+t))\]
where $T$ is the period of the applied periodic force and $t=1,2,\ldots T/2$. 
It is clear that $m(t)$ is just difference between number of times 
the value of the variable $x$ is greater than zero and that it is
less than zero, at times  $t$ modulo $T$ where  $ 1\leq t \leq T/2$.
We normalize $m(t)$ properly so that $r(t)=m(t)/M$ is confined between 1 and 
-1. ($M$ is the maximum value that $m(t)$ takes. By symmetry, we would expect 
it to be the same on positive and negative side.)
This gives one of the branches of hysteresis loop. The other
branch can be constructed by symmetry or computed by
\[m(T/2-t)={\frac{1}{N}}\sum_{j=1}^N sgn(x(jT+T/2+t))\] 
for $t=1,2,\ldots ,T /2$ 
and $r(t)=m(t)/M$.  Here one can have two
extreme cases. If the response is exactly in tune with 
the field, the magnetization will be zero at zero value of the periodic
force (no remnant magnetization) and the hysteresis area will be zero.
On the contrary, if the response is so late that only at the end of the 
half-cycle, the flips start occuring, loop will have maximum area.
Thus more delayed the response is, higher is the area of hysteresis
loop and hence is the popular notion of hysteresis area giving 
indication of losses in the system.
One would a priori think that as the noise intensity $D$
increases, which is an equivalent of increasing temperature, the hopping 
between the different states will be faciliated and thus the area of the loop 
will decrease at larger noise values. This is exactly what our observation 
is! We have plotted the area of hysteresis loop for the
system defined by eq.\ref{a)} in Fig 4b. Here,
we see a clear decrease in hysteresis loop area with increase
in noise intensity. As noise increases, the frictional losses are reduced 
since the system will not stay in metastable state for long. A sudden 
fluctuation will force it to respond to the
 signal and there will be little memory or remnant magnetization
in the system. Apart from the area of the hysteresis loop,
the shape of the hysteresis is also an interesting object to investigate.
Though for a noise higher than some critical value, 
the maxima and  minima in  magnetization start 
occurring at the maxima and minima of the field, for smaller noise they occur at
different times. (See, Fig. 4a, where we have plotted the hysteresis
loop for four different values of noise intensity.) We can see that
the response is delayed from the field by  a finite 
phase shift. This phase shift reduces with increasing noise\cite{fnn}. 
The results indicate that  in 
the limit of small noise the phase shift is  of the 
order of quarter of the total period, -$\pi/2$ which is expected\cite{Gamm}
for a two-state analysis that does  not 
take in account intra-well fluctuations.

\section{SR in Coupled Map Lattices}

Let us discuss cooperative phenomena possible in the spatially
extended versions of this system. Spatially discretized periodically
forced time dependent Ginzburg-Landau equation \cite{ben85}
, as well as one dimensional array of coupled bistable 
oscillators \cite{Lind}  have been studied before.
Major result in  the work by Lindner and coworkers 
\cite{Lind} has been that the largest value of $SNR$ is higher for a given
oscillator of the coupled system as compared to uncoupled one.
However,
the maximum of $SNR$ does not occur at the same value of noise intensity.
Not only the best value of $SNR$ is higher for
the coupled case, the value of $SNR$ at a given value of noise intesity
$D$ is better for the coupled system as compared to uncoupled one.
A simple interpretation that can be offered for
the above observation is that even when an oscillator
misses the interstate switching, the nearby oscillators may not,
thus forcing the individual oscillator to switch. 
This cooperative behavior should induce enhanced regularity 
in the switching of the oscillators and increase $SNR$ for a
given oscillator. 
The simple interpretation  
which can be true for any inter-state switching
mechanism should also
work for the systems we are studying.  

We define the following spatially extended system.
We will follow a evolution scheme popularly known as coupled map
lattice. Let us consider a linear array of length $N$. At each
lattice point $i$ ($i=1,\ldots N$) we attach a variable $x_i(t)$
at time $t$.
The time evolution of $x_i(t)$ is described by 
\begin{equation}
x_i(t+1)=(1-\epsilon) \; F(x_i(t))+{\frac{\epsilon}{2}}\;
         (F(x_{i-1}(t))+F(x_{i+1}(t))
\label{CML}
\end{equation}
for $2 \leq i \leq N-1$ and
\begin{eqnarray}
x_1(t+1)=(1-\epsilon) \; F(x_1(t))+\epsilon F(x_{2}(t)) \nonumber\\
x_N(t+1)=(1-\epsilon) \; F(x_N(t))+\epsilon F(x_{N-1}(t))
\end{eqnarray}
Where $F$ denotes some time evolving map. Of course,
we will be in particular interested in single maps that show
SR. (e.g eq.\ref{a)}). 

We have used the maps defined by eqs. \ref{a)},\ref{a1)},\ref{c)} and
\ref{c1)} as function $F$ in the above equation and have made a
detailed study of $SNR$ at various values of coupling $\epsilon$ and
noise intensity $D$ for $N=8$. Following Lindner {\it et al} \cite{Lind}
we have
looked at the response of the middle oscillator.  
Fig. 5a) and 5b) show the best value of $SNR$ as a function of
coupling for maps \ref{a)} and \ref{a1)}, i.e.,  the tanh map 
with additive and
parametric noise while Figs. 5c) and 5d) show the same for maps defined by
eq. \ref{c)} and eq.\ref{c1)}, {\it i.e.}, 
chaotic map with additive and multiplicative
noise  respectively.
We have the following observations.
a) In all these cases, the $SNR$ of the middle oscillator as a function of 
noise for a given value of coupling is 
nonmonotonic and shows a peak at some optimal value as in single
map case. The optimal value of noise need not be the same as one for
uncoupled case.
b) If the single system is 
perturbed with additive noise the coupling between 
such systems always enhances 
the $SNR$, i.e., best $SNR$ for the coupled
system is better than the one for a single oscillator case.
c) If the single system is perturbed by parametric noise, 
the coupling between such systems does not enhance $SNR$ much.
In fact for a large value of coupling, there is a clear 
decrease in $SNR$.

While observations a) and b) are in tune with the studies by
Lindner {\it et al} \cite{Lind} in coupled
bistable systems, the case with parametric noise has not 
been studied before.

We feel that the reason for this qualitative difference is following.
When two systems with different parameters are coupled,
i.e., a system with high Kramer's rate is coupled to the
one with low Kramer's rate,  the Kramer's rate for the 
coupled system is like that of the slower system \cite{Nei}.  
For higher coupling, this effect is more pronounced.
Thus, instead of an induced switching, one could have a slowed
down switching in presence of coupling. In this context,
we would like to point out a rather curious outcome of
our numerical investigation that the best $SNR$ has the least value  
around  $\epsilon =2/3$ in the case of  parametric noise. This
is the value at which all the maps in the neighbourhood
have equal weight in eq. \ref{CML} ($1-\epsilon=\epsilon/2=1/3$). 

\section{Discussion}

Due to computational simplicity of the system defined above,
and its ability to produce key features, 
the system defined above holds promise for carrying out work in
various directions with relative ease in future.
Here we would like to point out that using a coupled map like
formulation, Oono and Puri have been able to get a {\it quantitative}
agreement  in modeling phase ordering dynamics of Ising-type
systems\cite{Oono}. Miller and
Huse have also found that chaotic coupled maps with hamiltonian
symmetry show a phase transition with static and dynamic critical
exponents consistent to the Ising class\cite{Mill}. On the other hand,
Ising system 
have been reported to show SR in 1-d and 2-d\cite{IS1,IS2}. As we have pointed out,
recently there have been studies on coupled bistable oscillators in 1-d
which shows SR\cite{Lind}. 
Although one does not expect all the detailed behavior in
one model to carry over to the other, all these things do point towards 
a broader universality between
coupled bistable oscillators, coupled maps and Ising systems.
The investigations
on these lines should be useful in understanding SR in spatially
extended systems which are  relatively  unexplored but clearly important
from various points of view. Given the wide variety of physical situations that
Ising model is able to simulate, this similarity should not come as
a surprise. In fact, the similarity between coupled bistable oscillators
and Ising type systems has been pointed out before \cite{Jung}
While studies of 
the Ising system itself will be useful in analytic investigations,
investigations in the coupled map type systems will be computationally 
more efficient.
Studies in globally coupled maps\cite{Jung92} and  detection of
noise induced transitions in maps\cite{Gang}
could be carried out
in these systems. Though our preliminary numerical investigations do
not indicate any scaling as in \cite{Lind}, the behavior of
$SNR$ as a function of coupling $\epsilon$ and number of maps
$N$ could to be studied further and these investigations
are in progress. One could also investigate SR in  
2-dimensional coupled map lattice since higher dimensional spatially
extended systems are not investigated. One more important question
that needs to be addressed is the effect of multiplicative noise
and disorder in SR in spatially extended systems.

Authors have enjoyed discussions with Prof. N. Kumar(RRI). RR
would like to thank UGC for financial support and RRI for hospitality
while HS would like to thank the Indian 
Academy of Sci. for the support for the visit to JNCASR where the work was
carried out.

\centerline {\Large{Figure Captions}}
\begin{itemize}

\item[Fig. 1]  The  maps $f_S$, $g_a$ and $h_r$
are shown in as defined in  eqs. \ref{1}, \ref{2} and \ref{3}
are depicted in a) b) and c) respectively. 

\item[Fig. 2] The  $SNR$ as a function of $D$  for
systems defined by eqs.  \ref{a)},\ref{b)},\ref{c)} and \ref{c1)} are  shown 
in figures a), b), c), d) and e) respectively.
$\omega=1/8,\;S=2,\;,z=.5$ for a) and  b, $\omega=1/8,\;
r=1.4,a=12$ for d) and e), and $\omega=1/32,p_0=.01,z=.1$ for  c).
Noise is gaussian for case of unbounded map defined by eq. \ref{a)}
while is uniform on an unit interval in other cases.

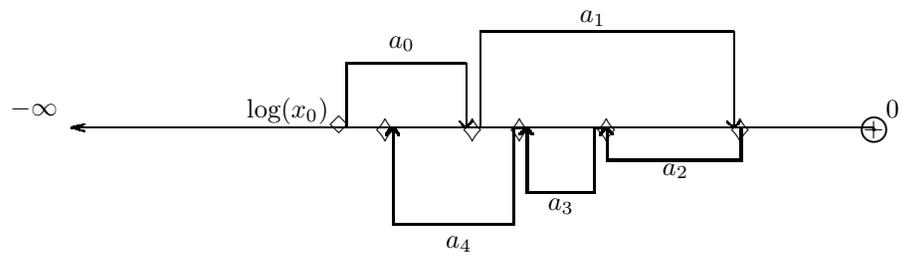
\item[Fig. 3] This figure shows the schematic diagram of how the
evolution under eq. \ref{b)} resembles a 1-d random walk on a logarithmic
scale with a  boundary condition at $ln(x)=0$ coming due to
modulo condition on the right hand side
while the
evolution is unbounded on the left hand side.

\item[Fig. 4] a)Hysteresis curve for $D=0.2$, $D=1 $, $D=1.4$ and $D=5.2$ for the map
defined by eq. \ref{a)} for $S=2$, $z=0.5$ and $T=1/\omega=32$. 
One can clearly see that for low $D$ maxima and minima of
magnetization are not in tune with the drive. 
b) Area of hysteresis loop as a function of noise intensity $D$ in this case.
One can clearly see a decrease at larger noise values.

\item[Fig. 5]  Best value of $SNR$ as a function of coupling $\epsilon$
for maps defined by eqs. \ref{a)}, \ref{a1)}, \ref{c)}, \ref{c1)}
are shown in a), b), c) and d). It is clear that while for additive noise the
best $SNR$ is enhanced, it is not so for parametric noise.

\end{itemize}

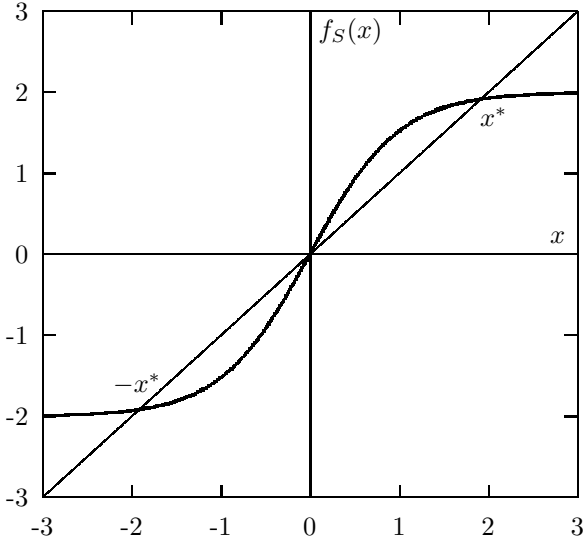
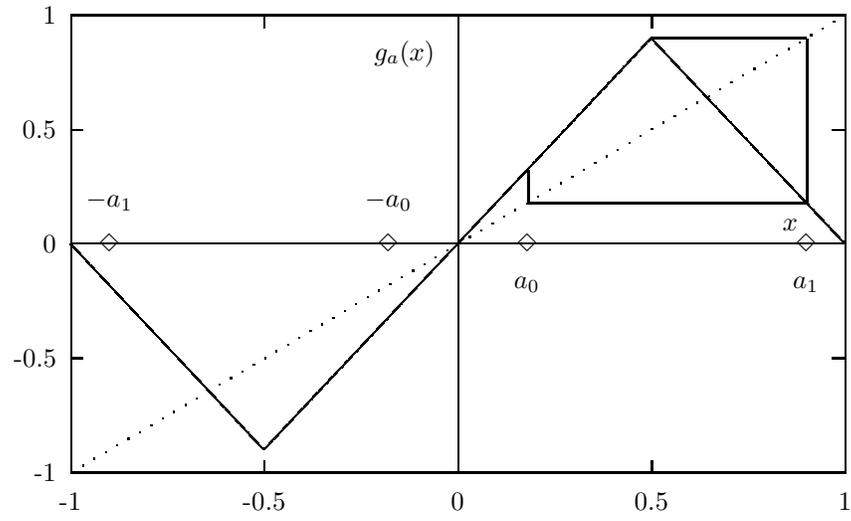
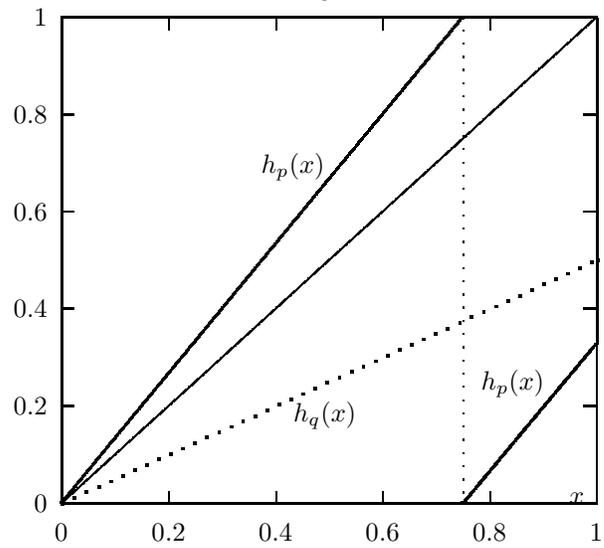
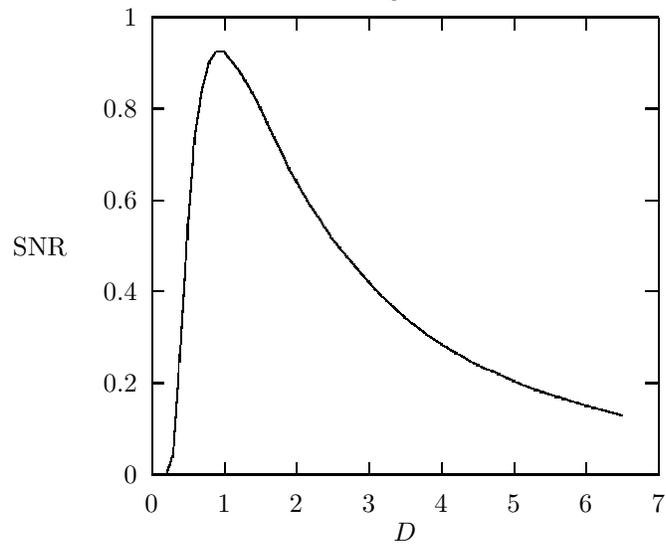
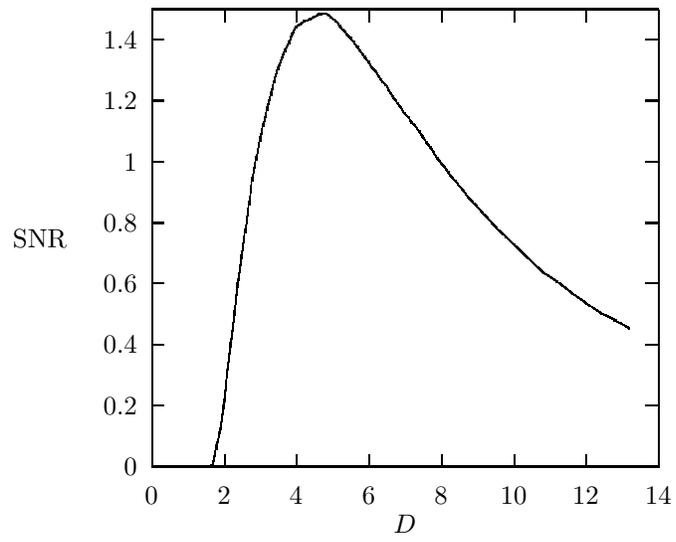
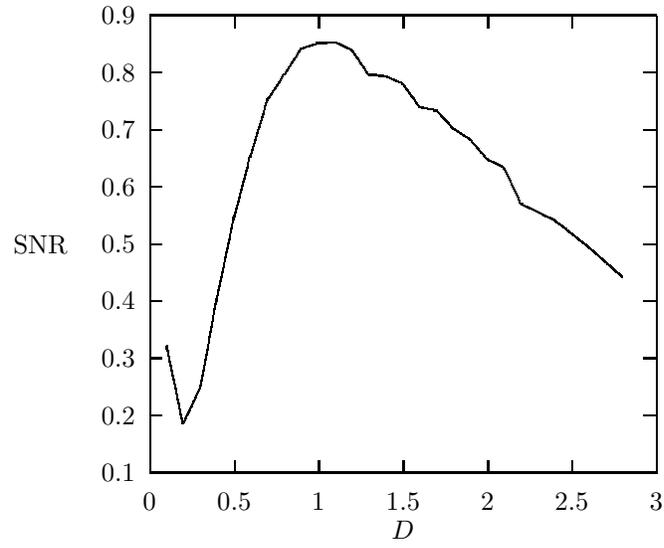
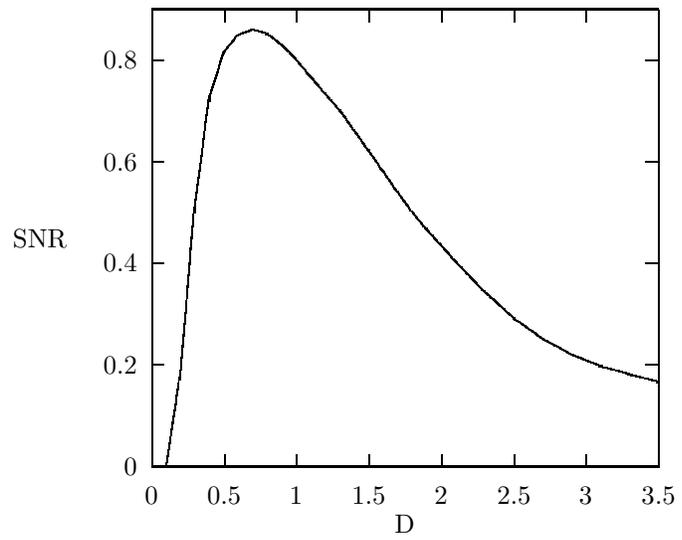
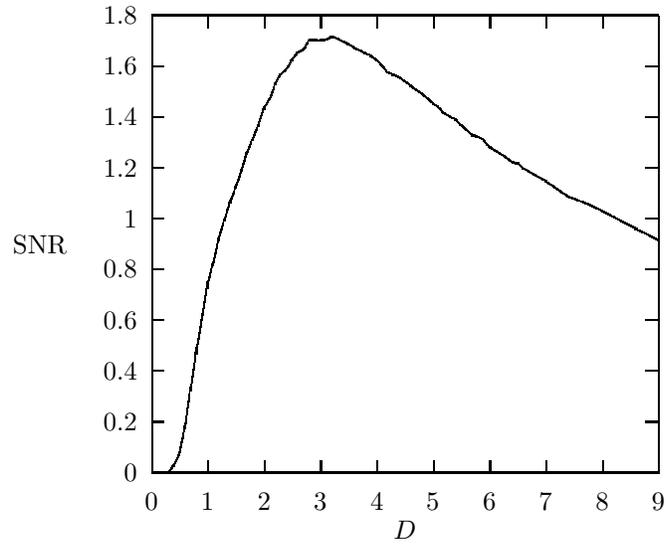
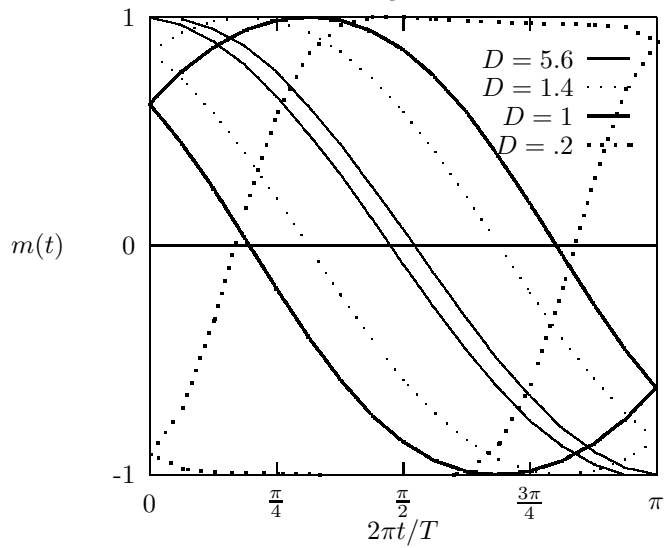
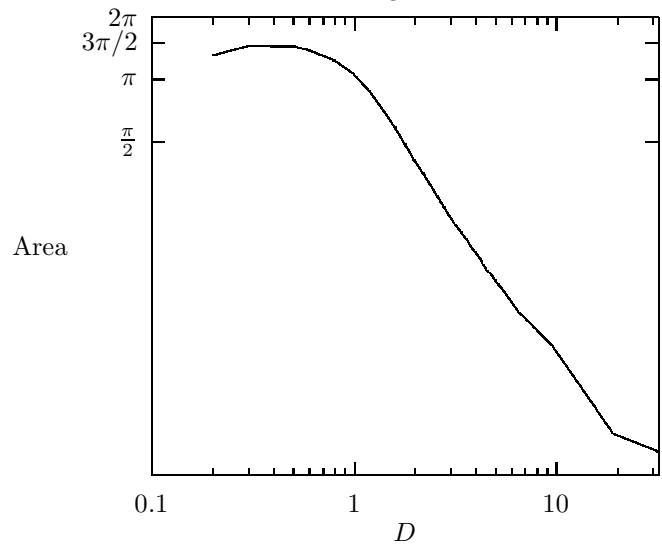
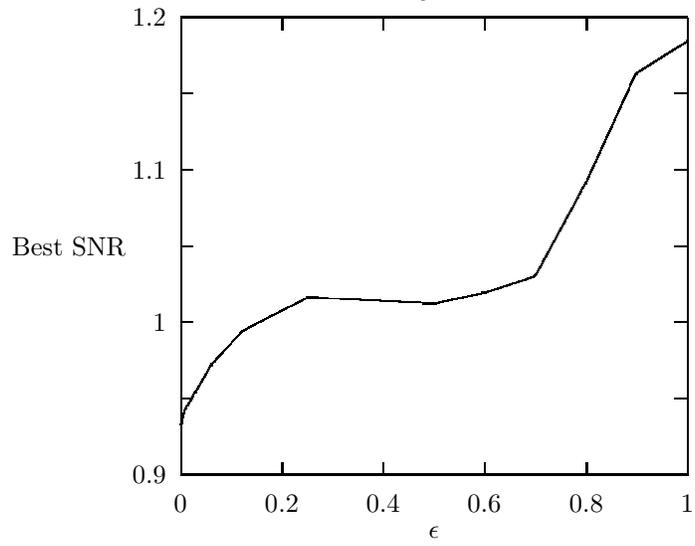
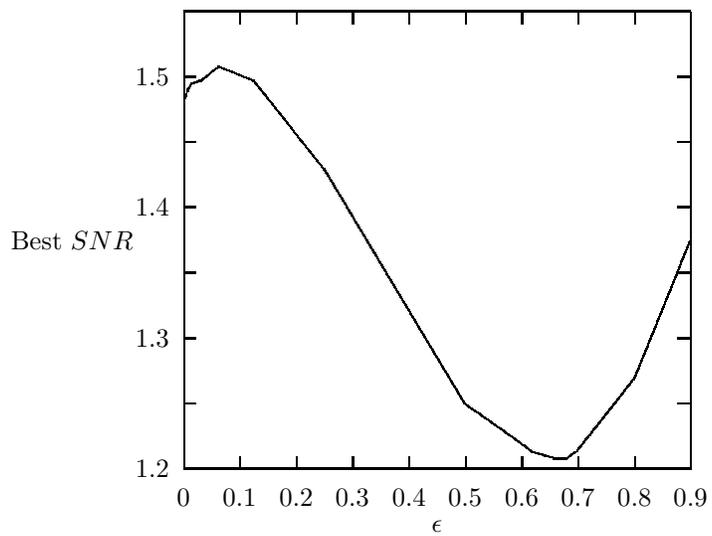
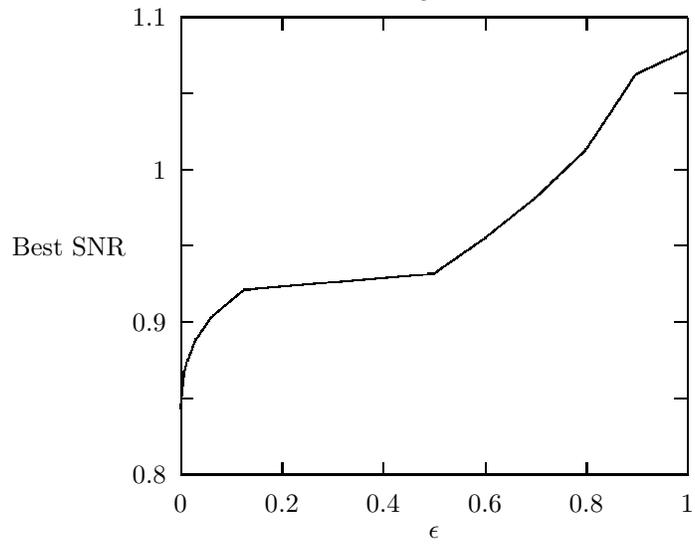
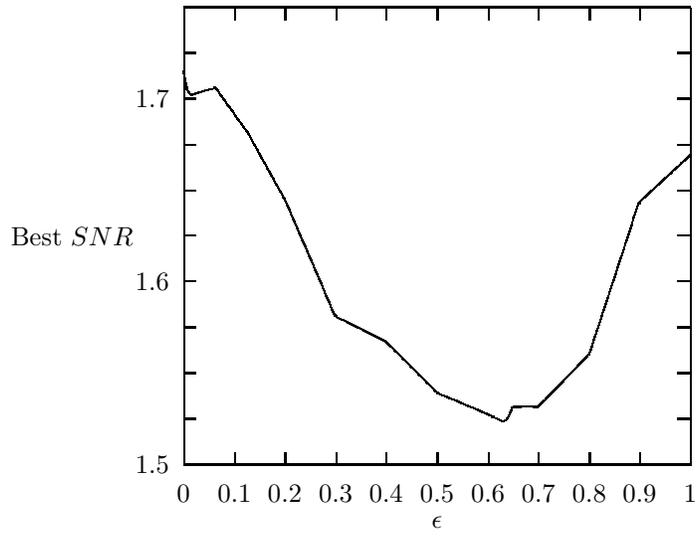
\end{document}